\documentclass{article}

\usepackage{arxiv}

\usepackage[utf8]{inputenc} % allow utf-8 input
\usepackage[T1]{fontenc}    % use 8-bit T1 fonts
\usepackage[hidelinks]{hyperref}       % hyperlinks
\usepackage{url}            % simple URL typesetting
\usepackage{booktabs}       % professional-quality tables
\usepackage{amsfonts}       % blackboard math symbols
\usepackage{nicefrac}       % compact symbols for 1/2, etc.
\usepackage{microtype}      % microtypography
\usepackage{lipsum}		% Can be removed after putting your text content
\usepackage{graphicx}
\usepackage{natbib}
\usepackage{doi}

\usepackage{setspace}
\title{The effect of estimating prevalences on the population-wise error rate}

\author{{Remi Luschei} \\
	Competence Center for Clinical Trials Bremen \\
	Institute for Statistics\\
	University of Bremen\\
	\texttt{rluschei@uni-bremen.de} \\
	\And
	{Werner Brannath} \\
	Competence Center for Clinical Trials Bremen\\
	Institute for Statistics \\
	University of Bremen\\
	\texttt{brannath@uni-bremen.de} \\
}

\date{\today}

\renewcommand{\headeright}{}
\renewcommand{\undertitle}{}
\hypersetup{
pdftitle={The effect of estimating prevalences on the PWER},
pdfauthor={Remi Luschei, Werner Brannath},
}

\usepackage{amsmath, amsfonts, mathtools, url, tikz, multirow, makecell}
\usepackage[titletoc, toc, page]{appendix}
\newcommand{\Pop}{\mathcal{P}}
\newcommand{\R}{\mathbb{R}}
\newcommand{\N}{\mathbb{N}}
\newcommand{\Var}{\text{Var}}
\newcommand{\Corr}{\text{Corr}}
\renewcommand{\P}{\mathbb{P}}
\newcommand{\PWER}{\text{PWER}}
\newcommand{\FWER}{\text{FWER}}
\newcommand{\SWER}{\text{SWER}}
\newcommand{\btheta}{\boldsymbol{\theta}}
\newcommand{\zero}{\boldsymbol{0}}
\newcommand{\bSigma}{\boldsymbol{\Sigma}}
\newcommand{\bpi}{\boldsymbol{\pi}}
\newtheorem{theorem}{Theorem}

\begin{document}
\maketitle

\begin{abstract}
The population-wise error rate (PWER) is a type I error rate for clinical trials with multiple target populations. In such trials, a treatment is tested for its efficacy in each population. The PWER is defined as the probability that a randomly selected, future patient will be exposed to an inefficient treatment based on the study results. It can be understood and computed as an average of strata-specific family wise error rates and involves the prevalences of these strata. A major issue of this concept is that the prevalences are usually unknown in practice, so that the PWER cannot be directly controlled. Instead, one could use an estimator based on the given sample, like their maximum-likelihood estimator under a multinomial distribution. In this article, we demonstrate through simulations that this does not substantially inflate the true PWER. We differentiate between the expected PWER, which is almost perfectly controlled, and study-specific values of the PWER which are conditioned on all subgroup sample sizes and vary within a narrow range. Thereby, we consider up to eight different overlapping populations and moderate to large sample sizes. In these settings, we also consider the maximum strata-wise family wise error rate, which is found to be, on average, at least bounded by twice the significance level used for PWER control.
\end{abstract}


\section{Introduction}

Many clinical trials in personalized medicine examine multiple hypotheses, each about the efficacy of a medical treatment in a specific patient population. This is done, for example, in umbrella trials that enroll patients with the same cancer type, but define multiple study arms based on different mutations of this cancer. Current examples of umbrella trials and challenges associated with these trials have been discussed by \cite{ouma}. One difficulty arises from the fact that in case of overlapping populations, taking a false test decision may affect more than one population. In this case, type I error control should be adjusted for multiplicity. For this purpose, \cite{brannath} introduced the population-wise error rate (PWER), a multiple type I error rate which is adapted to clinical trials with overlapping populations. It has been mentioned by \cite{ouma} and has the advantage of being more liberal than the family-wise error rate (FWER), while still controlling an average type I error. This allows to achieve higher power, what is particularly useful for examining rare diseases that often involve small sample sizes.

The PWER gives the probability that a randomly selected patient will receive an inefficient treatment. For calculating it, the overall population is partitioned into disjoint strata containing the patients that are affected by the same test decisions. For each stratum, the respective multiple type I error probability is then weighted by its relative prevalence, i.e.\ the proportion of the stratum in the total population. In practice, however, the prevalences are often unknown and must be estimated from the study sample. One way to do this, if no further knowledge about the prevalences is available, is to take the maximum-likelihood-estimator from the multinomial distribution. This article aims to examine whether the use of this estimator inflates the true PWER and to which extent. To this end, we will carry out simulations in which the relative prevalences will be estimated and used for the calculation of the rejection boundaries in order to control the estimated PWER at the pre-specified level. We will then compare the true PWER (using the true prevalences) to this significance level. 

Another important question regarding PWER control is to what extent the FWER can then still be controlled. In this context, the FWER corresponds to the maximal risk for future patients to be assigned to an inefficient treatment, while the PWER only controls the average risk. \cite{brannath} have specified several theoretical bounds for the FWER, which are somewhat rough and can often be improved by the simulated values. Specifically, we will see that the FWER is often less than twice as large as the significance level of PWER control.

We will also cover a special situation that may arise at the calculation of the estimated PWER, when no patients are recruited from (at least) one stratum. In this case, the concerned strata are neglected by the PWER, such that a randomly selected person from one of these strata could have a greater chance of receiving an inefficient treatment than the FWER level. Our solution is to introduce a minimal prevalence to include all concerned strata-wise error rates into the PWER. In our simulations we will therefore also examine to what extent PWER-controlling tests with a minimal prevalence actually become more conservative in this situation.
 
We start this article with the formal definition of the PWER in Section \ref{sec:pwer}. In Section 3, we construct the test statistics used to handle the population-wise testing problems. The presentation of the simulations and their results is done in Section \ref{sec:sim}, including a motivating example based on an umbrella trial data set from \cite{kesselmeier} in Section \ref{sec:ex}. The article ends with a discussion in Section \ref{sec:disc}. All simulations are done in R. The program files and outputs are available under \url{https://github.com/rluschei/PWER-Estimate-Prevalences}.

\section{The population-wise error rate (PWER)} \label{sec:pwer}

In this section we first recall the formal definition of the PWER from \cite{brannath} and present a result on least favorable parameter configurations for the PWER which will be used in the calculation of the rejection boundaries for PWER control.

\subsection{Definition}

Let $\Pop_1, \dots, \Pop_m$ be the sets representing the given patient populations. They may, in particular, be overlapping. For every $i \in I = \{1,\dots,m\}$ we want to test a treatment $T_i$ in the population $\Pop_i$. So we are interested in testing the null hypotheses $H_i \colon \theta_i \leq 0$, where $\theta_i = \theta(\Pop_i, T_i) \in \R$ denotes the effect of $T_i$ in comparison to a control treatment $C$ in the population $\Pop_i$. To define the PWER, we partition the overall population $\Pop = \cup_{i=1}^m\Pop_i$ into the disjoint strata $\Pop_J \coloneqq \left( \cap_{j \in J} \Pop_j \right) \backslash \left( \cup_{k \in I \setminus J} \Pop_k \right), J \subseteq I$. Each $\Pop_J$ includes all individuals affected by the treatments indexed in $J$. Figure \ref{fig:part} illustrates the resulting partition for $m=3$ overlapping populations. We denote the relative prevalence of $\Pop_J$ among $\Pop$ by $\pi_J$. Since the prevalences are usually unknown in practice, we will later propose how to estimate them. The \emph{population-wise error rate} of the multiple testing problem is defined by \[\PWER = \sum_{J \subseteq I}\pi_J \P(\text{falsely reject any $H_j$ with $j \in J$}).\] Hence, it gives the average probability of committing at least one type I error that would concern the individuals in $\Pop$. The PWER is more liberal than the family-wise-error-rate (FWER), i.e.\ it fulfils $\PWER \leq \FWER$ (see \cite{brannath}).

\begin{figure}[ht]
\centering
\begin{tikzpicture}
\draw (0,0) ellipse(2cm and 1.5cm);
\draw (-1,1.5) ellipse(2cm and 1.5cm);
\draw (1,1.5) ellipse(2cm and 1.5cm);
\node at (-2.75,2.75) {$\Pop_1$};
\node at (2.75, 2.75) {$\Pop_2$};
\node at (-1.75,1.75) {$\Pop_{\{1\}}$};
\node at (1.75,1.75) {$\Pop_{\{2\}}$};
\node at (0,-0.625) {$\Pop_{\{3\}}$};
\node at (0,2) {$\Pop_{\{1,2\}}$};
\node at (0,1) {$\Pop_{\{1,2,3\}}$};
\node at (0, -1.8) {$\Pop_3$};
\node at (-1.2,0.5) {$\Pop_{\{1,3\}}$};
\node at (1.25,0.5) {$\Pop_{\{2,3\}}$};
\end{tikzpicture} 
\caption{Partition of $m=3$ overlapping populations}
\label{fig:part}
\end{figure}

However, in order to control the PWER with estimated prevalences, we need to move to stronger null hypotheses. For any subset $J\subseteq I$ and $i \in J$, we define the null hypothesis $H_{J,T_i}: \theta_{J,T_i} \leq 0$, which states that treatment $T_i$ is not better than the control in the stratum $\Pop_J$. We will test the intersection hypotheses $H_i = \cap_{J \subseteq I: \, i \in J} H_{J, T_i}$ for each $i \in I$, stating that no treatment relevant to patients in $\Pop_i$ is more effective than the control. We assume that each $H_i$ is tested by a right-tailed test with the test statistic $Z_i$ and a critical value $c_i$. Then the PWER equals to \[\PWER_{\btheta} = \sum_{J\subseteq I}\,\pi_J\, \P_{\btheta}\left(\bigcup_{j\in J \cap I_0(\btheta)}\{Z_j > c_j\}\right),\] where the vector $\boldsymbol{\theta} = (\btheta_J)_{J \subseteq I}$ with $\btheta_J = (\theta_{J, T_i})_{i \in J}$ equals the true effects, and the true null hypotheses are indexed by $I_0(\boldsymbol{\theta}) = \{i \in I: H_i \text{ is true}\}$. In the following we will mostly consider equal critical values ($c_i = c$ for all $i \in I$) for the reasons given in \cite{brannath}.

\subsection{Least favorable parameter configurations} \label{sec:lfc}

Under two additional assumptions it can be shown that the null vector $\btheta=\zero$ is a worst case parameter for which the PWER attains its maximum. The first condition is the subset pivotality assumption (see \cite{westfall} and \cite{dickhaus}), which states that the multivariate distributions of the vectors $(Z_j)_{j \in J\cap I_0(\btheta)}$ do not depend on the truth or falsity of those hypotheses they are not relevant for. It is formally described by
\begin{align} \label{eq:sp}
\forall\, J \subseteq I, J \neq \emptyset: \forall\,\btheta \in H_J \, \exists\, \btheta^* \in H_I :  \P_{\btheta}^{\max_{j \in J}Z_j} = \P_{\btheta^*}^{\max_{j \in J}Z_j},
\end{align}
where $H_J \coloneqq \bigcap_{j \in J}H_j, J \subseteq I$, and $\P_{\btheta}^{\max_{j \in J}Z_j}$ denotes the distribution of $\max_{j \in J}Z_j$ in dependence of the parameter $\btheta$.
The second assumption is stochastic monotonicity of the subvectors $(Z_j)_{j\in J}$, stating that
\begin{align} \label{eq:sm}
\forall\,J \subseteq I, J \neq \emptyset: \forall\, \btheta_1, \btheta_2 \in H_I, \btheta_1 \leq \btheta_2: \P_{\btheta_1}\left(\max_{j \in J}Z_j > c \right) \leq \P_{\btheta_2}\left(\max_{j \in J}Z_j > c  \right).
\end{align}
Here the relation $\btheta_1 \leq \btheta_2$ is meant component-wise.

\begin{theorem} \label{thm1}
Under conditions (\ref{eq:sp}) and (\ref{eq:sm}) one obtains $\sup_{\btheta}\PWER_{\btheta} = \PWER_{\btheta^*}$ for $\btheta^* = \zero$.
\end{theorem}

The proof can be found in Appendix \ref{app:proof_thm1}. We call every $\btheta^*$ fulfilling the equation from Theorem \ref{thm1} a least favorable parameter configuration (LFC) for the PWER. If an LFC exists, the PWER can be strongly controlled at a given significance level $\alpha \in (0,1)$, by determining the smallest critical value $c$ that satisfies $\PWER_{\btheta^*}(c) \leq \alpha$. Note that this will also bound the FWER to a certain extent -- some theoretical bounds are given in \cite{brannath} and we will also investigate this numerically in Section \ref{sec:fwer}.

\section{Test statistics for a single-stage design} \label{sec:des}  

We now want to construct the vector of test statistics $\boldsymbol{Z}=(Z_i)_{i\in I}$ in order to conduct the multiple test. We follow the approach proposed by \cite{hillner} but replace the true prevalences used there by the given sample proportions. This allows for a concrete calculation  in practice, e.g.\ under an umbrella trial framework with a common control as described in \cite{ouma}.

For every $i \in I$, let $\mu_{i, T_i} \in \R$ be the expected response under treatment $T_i$ in the population $\Pop_i$ and let $\mu_{i,C} \in \R$ be the expected response under the control $C$ in $\Pop_i$. Then the mean effect difference in population $\Pop_i$ is $\theta_i = \mu_{i, T_i} - \mu_{i,C}$. Additionally, for every $J \subseteq I$ and $j \in J$ we define the expectations $\mu_{J,T_j} \in \R$ under $T_j$ and $\mu_{J,C} \in \R$ under $C$ in $\Pop_J$. The expectations in the population $\Pop_i$ are then \[\mu_{i,T_i} = \sum_{J \subseteq I:\,i \in J} \frac{\pi_J}{\pi_i} \mu_{J,T_i} \quad \text{and} \quad \mu_{i,C} = \sum_{J \subseteq I:\,i \in J} \frac{\pi_J}{\pi_i}\mu_{J,C} ,\] where $\pi_i$ denotes the sum of all $\pi_J$ with $i \in J$ (and equals the prevalence of $\Pop_i$). Suppose that a sample of $N$ patients is given, in which $n_J$ patients belong to the stratum $\Pop_J$, for every $J \subseteq I$. Let $n_{J,T}$ be the number of patients assigned to treatment $T$ in stratum $\Pop_J$, with $T \in \mathcal{T}_J \coloneqq \{T_j: j \in J\} \cup \{C\}$, and let $n_{i,T} \coloneqq \sum_{J \subseteq I: \, i \in J} n_{J, T}$ be the number of patients assigned to $T$ in population $\Pop_i$. In case of a stratified randomization, i.e.\ in every stratum $\Pop_J$ the patients are assigned to the treatments evenly, we have $n_{J,T} = n_J/(|J| + 1)$.

\subsection{Normal distribution model with known and heterogeneous variances} \label{sec:knownvar}

In every stratum $\Pop_J$, let us represent the measured responses under the treatment $T \in \mathcal{T}_J$ by the normally distributed random variables $ X_{J, T}^{(k)} \sim \text{N}(\mu_{J, T}, \sigma_{J, T}^2)$, for $k = 1, \dots, n_{J,T}$. The variances $\sigma_{J, T}^2 > 0$ are here assumed to be known, and all observations are assumed to be independent. For every $J$ and for every treatment $T \in \mathcal{T}_J$ we define the strata-wise arithmetic mean $\bar{X}_{J,T} \coloneqq  n_{J,T}^{-1} \sum_{k=1}^{n_{J,T}} X_{J, T}^{(k)}$. Then we can estimate the expected responses $\mu_{i,T_i}$ and $\mu_{i,C}$ through \[\hat{\mu}_{i,T_i} \coloneqq \sum_{J \subseteq I:\, i \in J}\frac{n_{J, T_i}}{n_{i,T_i}}\bar{X}_{J,T_i} \quad \text{and} \quad \hat{\mu}_{i,C} \coloneqq  \sum_{J \subseteq I: \, i \in J} \frac{n_{J, C}}{n_{i,C}}\bar{X}_{J,C}. \] This provides the test statistics \[Z_i \coloneqq \frac{\hat{\mu}_{i,T_i} - \hat{\mu}_{i,C}}{\sqrt{V_i}},  i \in I \] with \[V_i \coloneqq \Var(\hat{\mu}_{i,T_i} - \hat{\mu}_{i,C}) = \sum_{J \subseteq I: \, i \in J} \frac{n_{J, T_i}}{n_{i, T_i}^2} \sigma_{J, T_i}^2 + \frac{n_{J, C}}{n_{i, C}^2} \sigma_{J, C}^2.\]
Conditional on the sample sizes, the vector $\boldsymbol{Z} = (Z_i)_{i \in I}$ follows a multivariate normal distribution with location $\boldsymbol{\nu}= (\nu_i)_{i \in I}$ given by
\begin{align*}
\nu_i = \frac{1}{\sqrt{V_i}}\left(\sum_{J\subseteq I:\, i \in J}\frac{n_{J,T_i}}{n_{i,T_i}} \mu_{J,T_i} - \frac{n_{J,C}}{n_{i,C}} \mu_{J,C} \right), i \in I. 
\end{align*}
The correlation matrix $\boldsymbol{\Sigma} = (\Sigma_{ij})_{i,j \in I}$ takes a different form depending on whether and which treatments are the same. When all treatments are different ($T_i \neq T_j$ for $i \neq j$), we have 
\begin{align*}
\Sigma_{ij} = \frac{\sum_{J \subseteq I:\, i,j \in J} n_{J,C} \sigma^2_{J, C}}{n_{i,C} n_{j,C} \sqrt{V_iV_j}}, i \neq j.
\end{align*}
The derivation of the above results can be found in Appendix \ref{app:corr}.

The vector $\boldsymbol{Z}$ fulfills the conditions (\ref{eq:sp}) and (\ref{eq:sm}), so that $\btheta= \zero$ is an LFC for the PWER. Since $\boldsymbol{\nu}$ converges to zero under $\btheta=\zero$, the maximal PWER of the asymptotic test can be calculated by
\begin{align} \label{eq:pwer}
\PWER = \sum_{J \subseteq I} \pi_J \left(1- \Phi_{\zero, \bSigma_J}(c, \dots, c) \right),
\end{align}  
where $\Phi_{\zero, \bSigma_J}$ denotes the cdf of the normal distribution with parameters $\zero \in \R^{|J|}$ and $\bSigma_J \coloneqq \left(\Sigma_{ij}\right)_{i, j \in J}$. 

\subsection{Known and homogeneous variances} \label{sec:known-hom-var}

When the variances $\sigma_{J, T}^2$ are known and homogeneous ($\sigma_{J, T}^2 = \sigma^2$ for all $J \subseteq I, T \in \mathcal{T}_J$), the representation of the test statistics simplifies to \[Z_i = \frac{\hat{\mu}_{i,T_i} - \hat{\mu}_{i,C}}{\sigma \sqrt{H_i}} \quad \text{with} \quad H_i = n_{i, T_i}^{-1}+n_{i,C}^{-1} \] and the correlation matrix simplifies to 
\begin{align}\label{eq:corr}
\Sigma_{ij} = \frac{\sum_{J \subseteq I:\, i,j \in J} n_{J,C}}{n_{i,C} n_{j,C} \sqrt{H_iH_j}}.
\end{align}
and is then even independent from $\sigma$.

\subsection{Unknown and homogeneous variances} \label{sec:unknown-hom-var}

We now assume the variances to be unknown, but still homogeneous. For every $J \subseteq I$ and $T\in \mathcal{T}_J$ an unbiased estimator of $\sigma^2$ is given by \[ \hat{\sigma}_{J, T}^2 = \frac{1}{n_{J,T}-1}\sum_{k=1}^{n_{J,T}} \left(X^{(k)}_{J,T} - \bar{X}_{J,T}\right)^2.\] Let $s$ be the number of strata treatment combinations $(J, T)$ with $n_{J, T} > 1$. We use the pooled variance estimator \[\hat{\sigma}^2 = \frac{\sum_{J \subseteq I, \,T\in\mathcal{T}_J}(n_{J,T}-1)\hat{\sigma}_{J,T}^2}{N-s}\] to construct the test statistics \[T_i \coloneqq \frac{\hat{\mu}_{i,T_i} - \hat{\mu}_{i,C}}{\hat{\sigma}\sqrt{H_i}}, i \in I. \] The vector $\left(T_i \right)_{i \in I}$ then follows a multivariate $t$-distribution with the scale matrix from fomula \ref{eq:corr} and $df = N-s$ degrees of freedom (see \cite{kotz}). Hence, under an unknown and common residual variance the PWER is obtained by replacing $\Phi_{\zero,\bSigma_J}$ with the cdf of the multivariate $t$-distribution in formula \ref{eq:pwer}.

\subsection{Unknown and heterogeneous variances}

To account for unknown and heterogeneous variances across strata and treatments, we need to include the individual variance estimators $\hat{\sigma}_{J, T}^2$ in our test statistics: \[T_i^* \coloneqq \frac{\hat{\mu}_{i,T_i} - \hat{\mu}_{i,C}}{\sqrt{\hat{V}_i}} \quad \text{with} \quad \hat{V}_i = \sum_{J \subseteq I:  \ i \in J} \frac{n_{J, T_i}}{n_{i, T_i}^2} \hat{\sigma}_{J, T_i}^2 + \frac{n_{J, C}}{n_{i, C}^2} \hat{\sigma}_{J, C}^2.\]  However, finding their joint distribution is an unsolved problem, so that the PWER cannot be calculated in this case. \cite{hasler} show that FWER-control can still be reached by computing individual critical values $c_i$ from a $t$-approximation with degrees of freedom according to \cite{satterthwaite} and plug-in estimation of the correlation matrix. In Section \ref{sec:unkown_het_var} we will show how this can be applied to PWER-control.

\section{Estimation of the population prevalences} \label{sec:sim} 

The populations $\Pop_1, \dots, \Pop_m$ are usually defined by certain inclusion and exclusion criteria and are assumed to have infinite size, because they not only include the study patients, but also patients outside of the trial and potential future patients. Consequently, the exact values of the relative prevalences $\pi_J$ are typically not known in practice and must be estimated in some way to compute the PWER. For this purpose, one could possibly use findings from previous studies. Another possibility is to utilize the sample sizes collected in the study. In the latter case, the vector $\boldsymbol{\pi} = (\pi_J)_{J \subseteq I}$ may be estimated by the maximum-likelihood-estimator (MLE) of the multinomial distribution with parameters $N$ and $\boldsymbol{\pi}$, whose discrete density  
$f \colon (n_J)_{J \subseteq I}  \mapsto 
N!  \prod_{J \subseteq I} \pi_J^{n_J}/n_J!$
gives the probability of selecting $n_J$ patients from the population $\Pop_J$ in a random sample of $N = \sum_{J \subseteq I}n_J$ patients. The MLE $\boldsymbol{\hat{\pi}} = (\hat{\pi}_J)_{J \subseteq I}$ has the form $\hat{\pi}_J = n_J/N$, and is a mean unbiased and asymptotically consistent estimator. This implies that for a given fixed critical value $c$ the PWER can be estimated mean unbiasedly and consistently by plugging in $\hat{\pi}_J$. However, in practice we would rather determine the  critical value such that the estimated PWER equals the significance level $\alpha$.

\subsection{Umbrella trial example} \label{sec:ex}

\begin{figure}[h]
\centering
\includegraphics[width=1\textwidth]{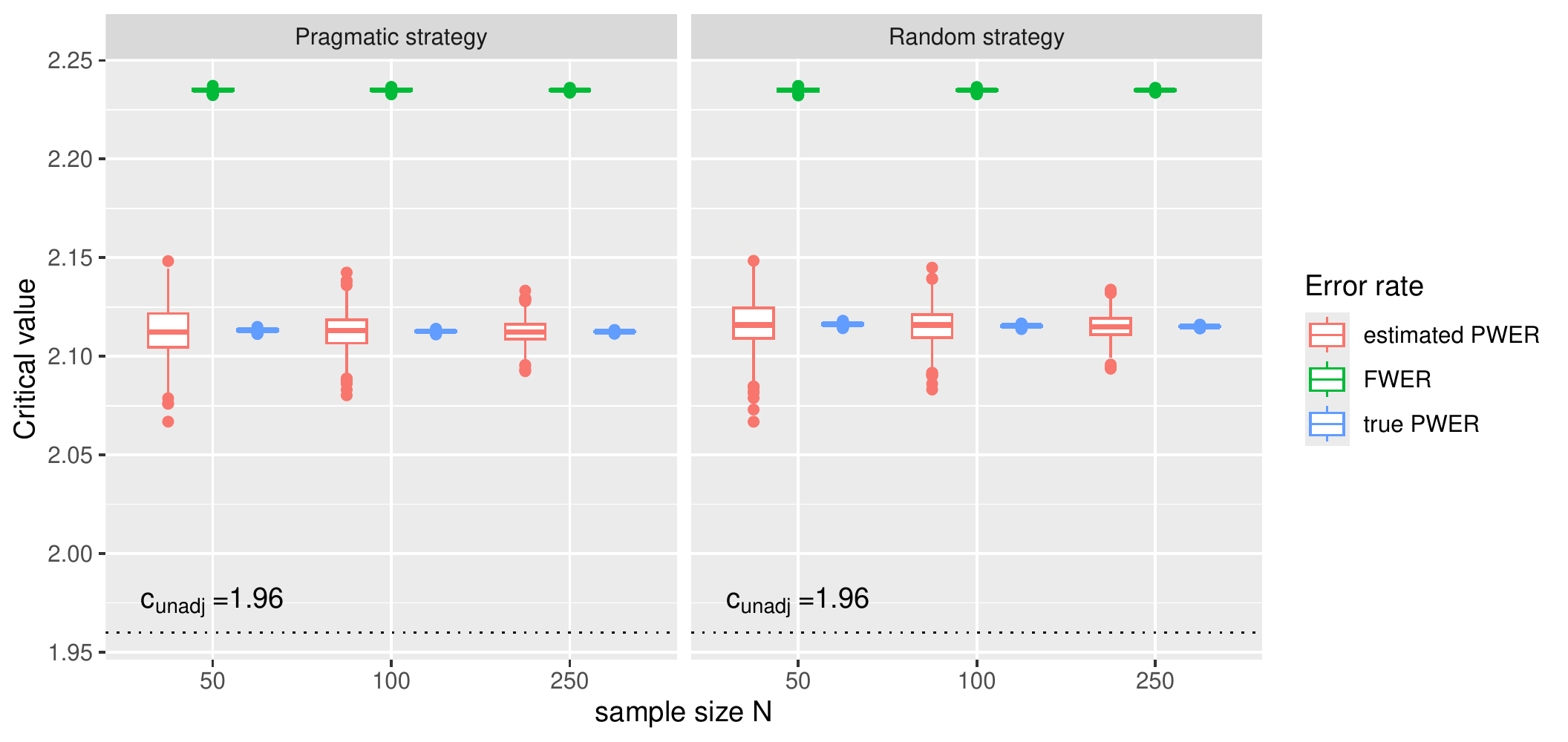}
\caption{Critical boundaries for the real data example from \cite{kesselmeier} under FWER-control, control of the estimated PWER and control of the true PWER. For each bootstrap sample, we utilize the allocation numbers and prevalence estimates that result from applying the pragmatic strategy and the random allocation strategy to calculate the FWER and the PWER. We obtain the true critical values by plugging the mean prevalence estimates into the PWER. We assume known and equal residual variances and use $\alpha = 0.025$ as significance level. The critical value for unadjusted testing is $c_\text{unadj} =\Phi^{-1}(1-\alpha)= 1.96$. FWER: family wise error rate; PWER: population-wise error rate.}
\label{fig:ex}
\end{figure}

To illustrate how the test decisions change under control of the estimated PWER compared to FWER control and unadjusted testing, we regard a real data example from \cite{kesselmeier}. It is introduced to investigate two treatment allocation strategies for patients that are eligible for multiple arms in umbrella trials: the pragmatic strategy assigning the patients to the eligible subtrial with currently fewest patients, and random allocation of these patients. The example is based on the MAXSEP study (\cite{brunkhorst}) that compared the effect of meropenem to the effect of a combination therapy with moxifloxacin and meropenem in patients with severe sepsis. \cite{kesselmeier} define two overlapping populations,
\begin{align*}
\Pop_1 &= \text{patients with baseline lactate value $> 2 \text{ mmol}/\text{L} $} \\
\Pop_2 &= \text{patients with baseline C-reactive proteine value $> 128 \text{ mg} / \text{L}$}
\end{align*}
and for these they build 1\,000 bootstrap samples of sizes $N \in \{100,200,500\}$ from the study data. For each they apply both allocation stategies, and also the gold-standard independent trial design (where the subtrials screen their patients independently for just one of the biomarkers). We use the resulting allocation numbers and prevalence estimates to compute the rejection boundaries one would get under FWER-control and control of the estimated PWER, for both allocation strategies. We assume the significance level $\alpha = 0.025$ and equal residual variances. In the independent subtrials case, one would typically not adjust for multiplicity, such that one would use $\Phi^{-1}(1-\alpha) = 1.96$ as critical value, where $\Phi^{-1}$ denotes the quantile function of the standard normal distribution. The distribution of the resulting critical values is plotted in figure \ref{fig:ex}. We see that they can be significantly reduced by replacing FWER control with PWER control and that PWER control leads to a good compromise between unadjusted testing and the more strict FWER control. In figure \ref{fig:ex} we also added the true critical boundaries for PWER control that we obtain from the true prevalences, to examine how well they are approximated by the estimated ones. Therefore we assume that the true prevalences are equal to the mean estimates over all the 1\,000 bootstrap samples. We see that the estimated values are close to the true ones (especially their average) and that the approximation improves with increasing sample size.

\subsection{Setup of the simulations} \label{sec:setup}

We now want to investigate the accuracy of the PWER estimation more systematically in some simulations. We will first focus on the cases where the distribution of the test statistics is known (Sections \ref{sec:knownvar} to \ref{sec:known-hom-var}) and will deal with the case of unknown, homogeneous variances in Section \ref{sec:unkown_het_var}. We find the estimated critical value $\hat{c}$ from the condition
\begin{align} \label{eq:pwer-est}
\widehat{\PWER}\left(\hat{c}\right) \coloneqq \sum_{J \subseteq I} \hat{\pi}_J \left(1- F_{\zero, \bSigma_J}(\hat{c}, \dots, \hat{c}) \right) = \alpha,
\end{align}
where $F_{\zero, \bSigma_J}$ denotes the cdf of the Gaussian or $t$-distribution. Our goal is now to find out if the true PWER 
\begin{align} \label{eq:true-pwer}
\PWER\left(\hat{c}\right) = \sum_{J \subseteq I} \pi_J \left(1- F_{\zero, \bSigma_J}(\hat{c}, \dots, \hat{c}) \right)
\end{align} 
is then still controlled at the significance level $\alpha$. Asymptotically, this is the case due to the almost sure convergence of $\hat{\pi}_J$ to $\pi_J$. We can also show that the sequence of critical values ($\hat{c}_N)$ converges in probability towards the true critical value $c(\bpi)$ that would be obtained from the true prevalences and correspondingly proportioned sample sizes. This is not even necessary for the proof of asymptotic PWER control, but could be of interest for further considerations (see the discussion). Both results are shown in Appendix \ref{app:conv}.

For our simulations we define the populations $\Pop_1, \dots, \Pop_m$ by $m$ different binary  biomarkers that are expressed with probabilities $p_1, \dots, p_m$. For every $i \in I$, the population $\Pop_i$ consists of all patients with the $i$-th biomarker being expressed. In particular, since the biomarkers are not assumed to be exclusive, these populations overlap. It should be noted that we do not consider the configuration where none of the biomarkers are present, since these patients would typically be excluded from the study for ethical reasons. Hence, we consider here the overall population of patients where at least one biomarker is expressed. In the first step of the simulation, we define the biomarker expression probabilities $p_i$ (for example by randomly generating them from the uniform distribution). From the $p_i$ we generate in turn a multinomial random vector containing the $2^m-1$ strata-wise sample sizes. We find $\hat{\boldsymbol{\pi}}$ by dividing this vector with the total sample size $N$. The critical value $\hat{c}$ is then computed from formula (\ref{eq:pwer-est}) and plugged into the true PWER (formula \ref{eq:true-pwer}). We repeat this procedure 10\,000 times, draw a boxplot of the resulting values of $\PWER(\hat{c})$, and tabulate their summary statistics.

\subsection{Results} \label{sec:results}

\begin{figure}[h]
\centering
\includegraphics[width=0.75\textwidth]{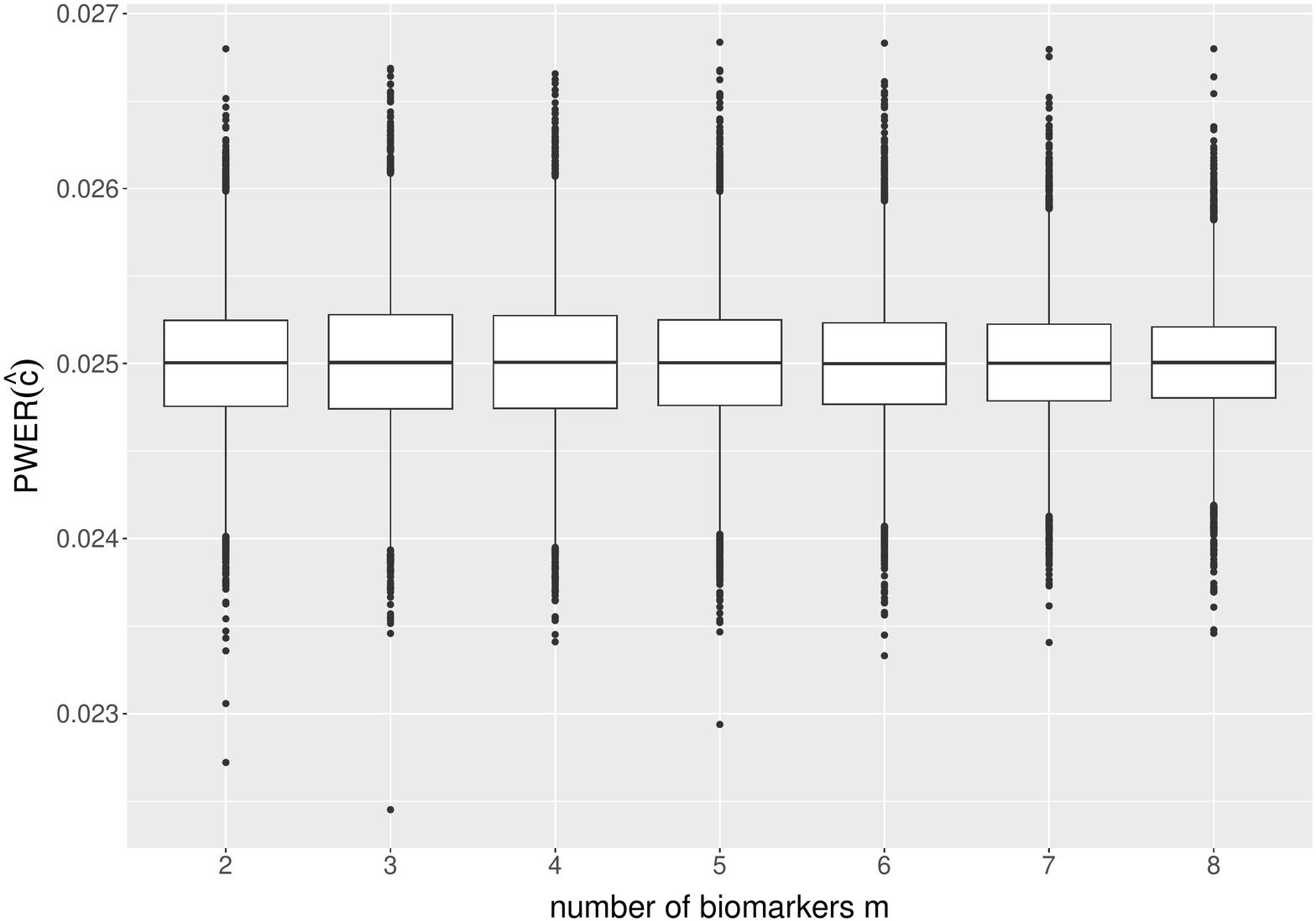}
\caption{Distribution of the true population-wise error rate (PWER) for different numbers $m$ of binary and independent biomarkers. This corresponds to $m$ overlapping populations with $2^m-1$ strata. The critical value $\hat{c}$ is computed from the estimated PWER, under the significance level $\alpha=0.025$. We assume a total sample size of $N=500$ screened patients and a stratified randomization to the treatments.}
\label{fig:sim1}
\end{figure}

We run the simulations described above for many different scenarios in which the following aspects are varied: 
\begin{itemize}
\item We consider different numbers $m$ of biomarkers, from $m=2$ to $m=8$ biomarkers. 
\item We consider different total sample sizes, from $N=25$ to $N=500$ patients.
\item We consider independent and dependent biomarkers.
\item We consider known and unknown homogeneous residual variances and known heterogeneous variances (see Section \ref{sec:unkown_het_var} for the unknown, heterogeneous case).
\item We consider pairwise different treatments ($T_i \neq T_j$ for $i \neq j$) and equal treatments ($T_i = T_j$).
\item We consider fixed biomarker expression probabilities (= fixed true prevalences) and probabilities that vary with each simulation run. 
\item We consider equal patient allocation to the treatments within the strata and random allocation.
\item We consider different significance levels ($\alpha = 0.025$ and $\alpha = 0.01$).
\end{itemize}
For all these configurations, the distribution of true PWER (obtained from formula \ref{eq:true-pwer}) is found to be very similar. The detailed results can be found in our Github repository. Exemplary, we now describe the results for the following configuration: We fix the total sample size at $N=500$ patients, which corresponds to a medium to large size for a multi-population study. We use $t$-distributed test statistics with the correlation matrix from formula (\ref{eq:corr}), and regard $m=2$ to $m=8$ independent biomarkers. The biomarker expression probabilities are randomly and independently generated from the uniform distribution on $(0,1)$ in each simulation run. With this we cover situations with up to $2^8-1=255$ different strata. Additionally, we assume equal allocation of the patients to the eligible treatments within the strata. As significance level we take $\alpha = 0.025$. The distribution of the true PWER for this configuration is plotted in figure \ref{fig:sim1}. One can see that for all $m$, the values are clustered quite tightly and symmetrically around $\alpha$. The mean values, which approximate the expected, overall PWER, are all equal to $\alpha$ at least up to the fourth decimal place, and the standard deviations are all smaller than $4.2 \cdot 10^{-4}$. The detailed summary statistics for figure \ref{fig:sim1} can be found in Appendix \ref{app:tables}. For these reasons, the true PWER appears to be well under control. Small variations may only occur in individual situations: 5.47 percent of the values in figure \ref{fig:sim1} are outside the interval $(0.02375, 0.02625)$ and thus deviate from $\alpha$ by at least 5 percent. In summary, this implies that, conditional on the actually observed sample sizes, the PWER may (with a small probability) be moderately inflated or deflated, however, is well under control in the average.

\begin{figure}[h]
\centering
\includegraphics[width=0.75\textwidth]{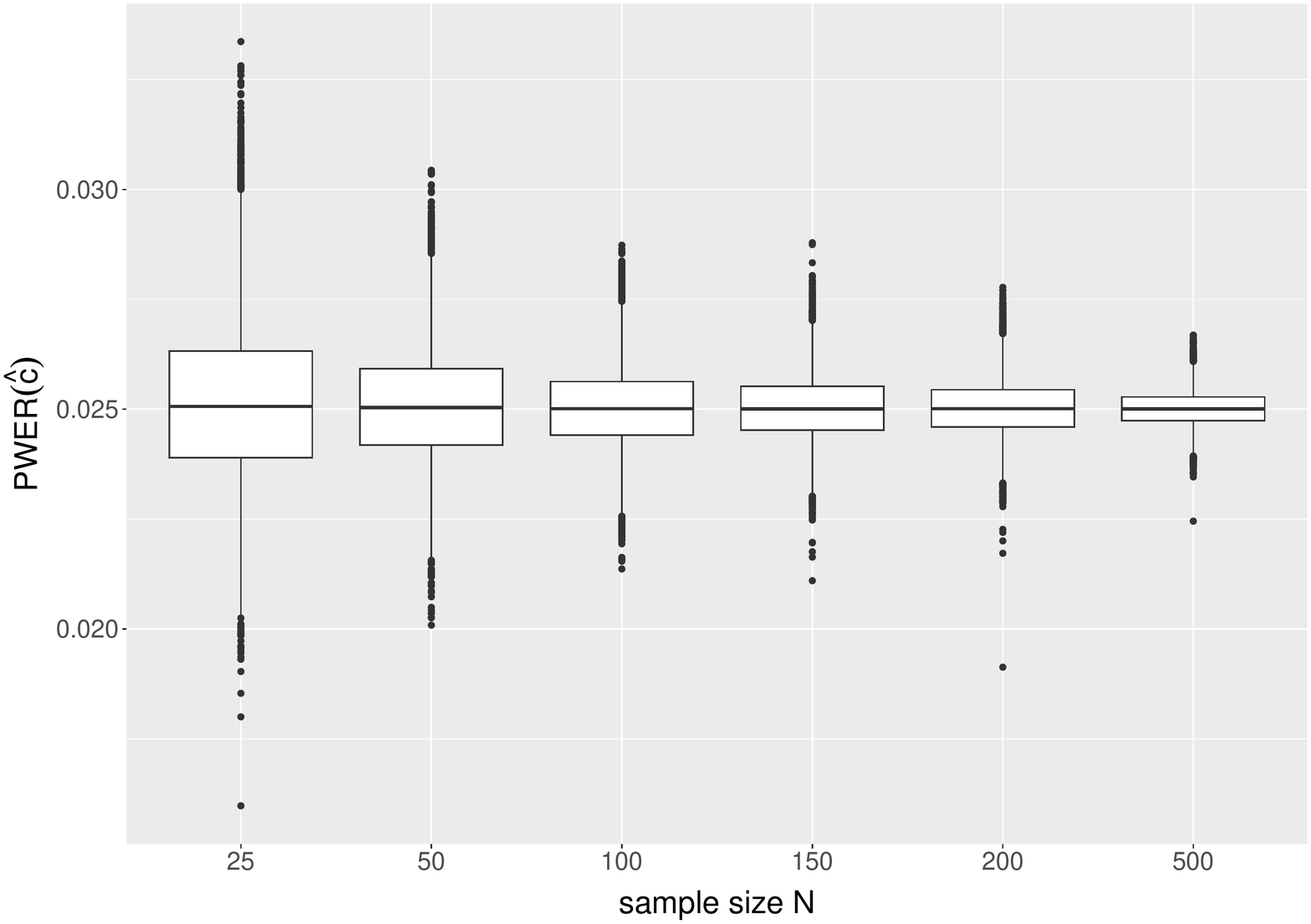}
\caption{Distribution of the true population-wise error rate (PWER) for different overall sample sizes $N$ and $m=3$ independent biomarkers. The critical value $\hat{c}$ is computed from the estimated PWER, under the significance level $\alpha=0.025$.}
\label{fig:sim2}
\end{figure}

Figure \ref{fig:sim2} shows the results for another configuration where the number of biomarkers is fixed at $m=3$ and the sample size varies from $N=25$ to $N=500$. While larger inaccuracies can occur at sample sizes $N=25$ or $N=50$, for $N=100$ all values of the true PWER already deviate less than 15 percent from the significance level, and for $N=200$ less than 10 percent. So the mentioned convergence of $\PWER(\hat{c})$ with respect to $N$ takes place very quickly and applies to practical situations.

In figure \ref{fig:sim1}, we see that the boxplot ranges decrease as the number $m$ of populations increases. This is probably due to the fastly increasing number of strata and thus decreasing prevalences, so that for larger $m$, single estimation inaccuracies in the strata probabilities do not have a major impact on the PWER. In practice, of course, the sample size would have to be increased in order to have enough patients in all strata. We get similar results when we assume equal prevalences for all strata and leave them constant over all simulation runs. However, if individual prevalences remain constant with an increasing number of strata, this is not the case any more. This can be seen, for example, when one prevalence is set to 0.5 independently of the number of strata. The range of the simulated values then increases with increasing $m$, up to a standard deviation of $7.7 \cdot 10^{-4}$ for $m=8$.

The assumption of independent biomarkers is not necessarily guaranteed in practice. But the above simulations can easily be extended to correlated biomarkers by deriving the biomarker probabilities from a normally distributed random vector with an arbitrary covariance matrix. We have done this for various cases and found no particular differences to the results presented above. The same applies to the other cases presented before.

\subsection{Marginal sum estimator} \label{sec:marg}

When defining the populations for our simulation, we assumed that all patients without any biomaker being expressed would not be included in the study. However, these patients usually also go through the screening process, so their number is usually known. We can use this to define another estimate. Let $\hat{\tau}_J, J \subseteq I$ denote the prevalence estimators that we get when including this stratum into the total population (so that we have $\hat{\tau}_{\emptyset} > 0$). In the independent biomarkers case, another possibility of estimating $\pi_J$ is then using the marginal prevalence estimator
\begin{align} \label{eq:marg-est}
\tilde{\pi}_J =  \left(\prod_{j \in J} \hat{p}_j \right)\left( \prod_{k \in I\setminus J} (1-\hat{p}_k)\right) / \left(1- \hat{\tau}_\emptyset \right),
\end{align}
which is based on the marginal frequencies $\hat{p}_j = \sum_{J \subseteq I:\, j \in J} \hat{\tau}_J$ of the strata. Under the independence assumption, this also gives consistent estimates. We adopted the marginal prevalence estimator into our simulations and got similar results to Section \ref{sec:results}: The deviation of the true PWER from $\alpha$ is slightly lower than before, with standard deviations lying between $2.7 \cdot10^{-4}$ ($m=2$) and $3.6 \cdot 10^{-4}$ ($m=4,5$).

\subsection{Behavior of the strata-wise FWER} \label{sec:fwer}

\begin{figure}[h] 
\centering
\includegraphics[width=\textwidth]{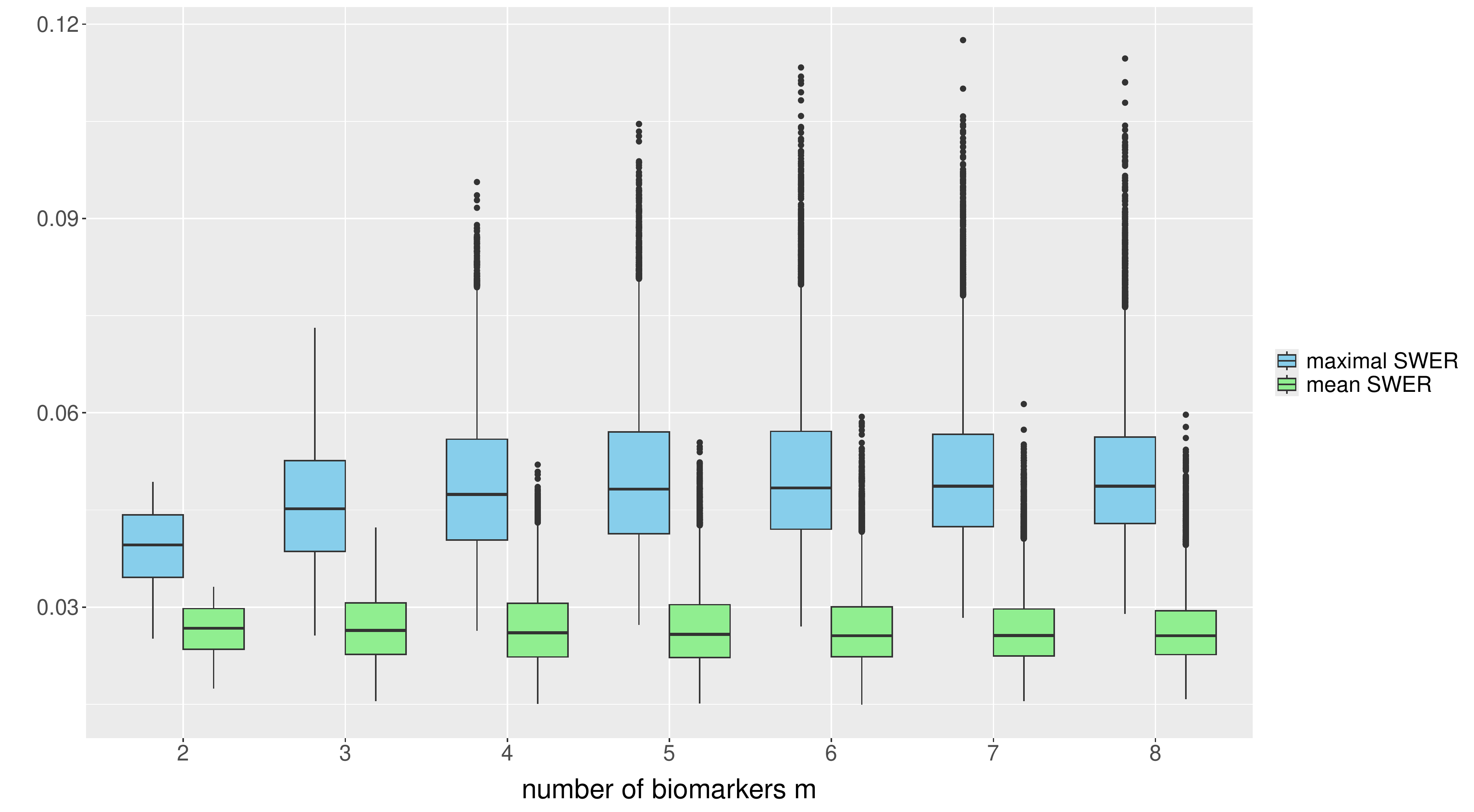}
\caption{Distribution of the the maximal strata-wise FWER and of the mean strata wise FWER for different numbers of biomarkers $m$, under PWER control at level $\alpha=0.025$. The overall sample size is fixed at $N=500$. FWER: family wise error rate; PWER: populationwise
error rate.}
\label{fig:sim3}
\end{figure}

In PWER-controlling test procedures, we may also be interested in the behavior of the single strata-wise family-wise error rates, which are defined by $\SWER_J(c)\coloneqq \P(\max_{j \in J} Z_j > c)$ for every $J \subseteq I, \Pop_J \neq \emptyset$, to verify whether excessive type I error probabilities occur in individual strata. \cite{brannath} give some upper bounds for $\SWER_J(c)$, whose quality depend on different factors like the prevalence $\pi_J$ or the number of biomarkers the stratum $\Pop_J$ belongs to. Under the setup of the previous simulations, we also examine the values of the maximum $\max_{J \subseteq I, \Pop_J \neq \emptyset} \SWER_J(\hat{c})$ (which equals the global FWER when $\Pop_I \neq \emptyset$) and of the mean over all $\SWER_J(\hat{c}), J \subseteq I, \Pop_J \neq \emptyset$. They are plotted in figure \ref{fig:sim3}, for the first case presented (with randomly generated true prevalences, independent biomarkers and unknown residual variances). We see in these simulations that on average, the maximal strata wise FWER is limited by $0.05 = 2 \alpha$. The mean SWER is on average only slightly larger than $\alpha$. We get very similar results in most other cases (see the detailed results in the repository). But we note that the bound $2\alpha$ for the maximal SWER is just an empirical finding and may be exceeded in individual cases (for specific randomly drawn true prevalences and strata-wise sample sizes) which can occur in individual studies. For example, even the average maximal strata-wise FWER is found to be much higher (up to 0.07455 for $m=8$ populations) in the case where we set one prevalence to 0.5, independently of the number of strata.

\subsection{Introduction of a minimal prevalence for neglected strata} \label{sec:min}

For very small but non-empty population strata $\Pop_J$, it may happen that no patients are sampled, meaning that they are not included in the estimated PWER. Then we would not directly account for the multiplicity in these strata and might therefore expose future patients to an increased risk of receiving inefficacious treatments. This is especially a problem for strata that are intersections of many different populations, since smaller intersections may partially be controlled by the larger ones. If the biomarkers are independent, a solution could be to replace the MLE of the prevalences with the marginal estimator $\tilde{\pi}_J$ from Section \ref{sec:marg}. Since these estimates are based on the empirical marginal prevalences of the larger subpopulations $\Pop_j$, the problem of missed prevalences is avoided or at least much reduced. In the general case, one could introduce a minimal prevalence $\pi_\text{min}$ to include all $\SWER_J$ with $n_J=0$ in the estimated PWER, and reduce the other prevalences proportionally. However, all strata with estimated prevalences smaller than $\pi_\text{min}$ should not be penalized, so we suggest to increase their weights to $\pi_\text{min}$ as well. The minimal prevalence should not be chosen too large, as this could result in greater inaccuracy of the estimations. One possible choice is $\pi_\text{min} = 1/(2^{m+1}-2)$, i.e.\ half of the prevalence that each stratum would have if all strata were of same size. Given the very small value of $\pi_\text{min}$, especially for high numbers of biomarkers, we are then unlikely to overestimate the true prevalences of the strata with no observations (and underestimate the others).

\begin{figure}[h] 
\centering
\includegraphics[width=\textwidth]{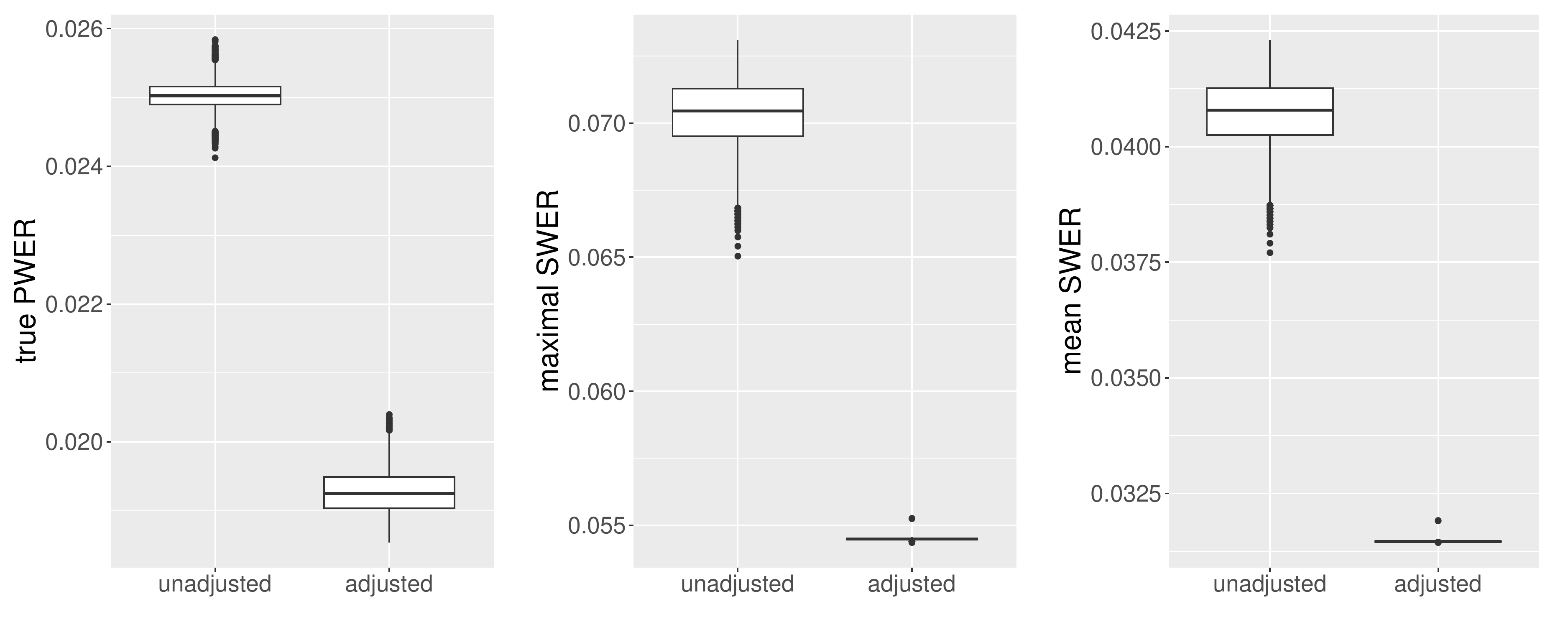}
\caption{Comparison of the true PWER (left plot), the maximal strata-wise FWER (middle plot) and the mean strata-wise FWER (right plot) with and without the adjustment by the minimal prevalence $\pi_\text{min} = 1/(2^{m+1}-2)$. We consider a setting with $m=3$ popoulations, $N=500$ patients and $\alpha=0.025$ as significance level for PWER-control. Only  cases in which at least one stratum has no observations are included here. PWER: population-wise error rate; FWER: family wise error rate.}
\label{fig4}
\end{figure}

This approach actually leads to more conservative tests: In additional simulations we restrict the biomarker probabilities $p_1$, ..., $p_m$ to the interval $(0,0.1)$, in order to achieve that large intersections preferably get small prevalences. We then compute the critical boundary $\hat{c}$ and the adjusted $\hat{c}_\text{min}$ resulting from weighting up all neglected strata by $\pi_\text{min}$. Figure \ref{fig4} compares the distributions of the true PWER, the maximal SWER and the mean SWER for these two boundaries, in a setting with $m=3$ populations. We see that in the unadjusted case, the maximal SWER is increased in comparison to the previous simulation results (because we only consider unfavorable cases here), and is reduced to an acceptable level by the suggested adjustment of the prevalences. The true PWER and the mean SWER are also becoming more conservative. For larger numbers of biomarkers $m$, this reduction becomes larger in absolute terms. For example, for $m=6$ the mean true PWER decreases from 0.025 to 0.01403 and the mean maximal SWER decreases from 0.12579 to 0.0723 by replacing $\hat{c}$ with $\hat{c}_\text{min}$. Note that in general $\hat{c}_\text{min}$ is not necessarily larger than $\hat{c}$, especially when strata belonging to only few populations have small prevalences. Therefore we would suggest comparing $\hat{c}$ and $\hat{c}_\text{min}$ and choosing the greater one for PWER control.

\subsection{The case of unknown, heterogeneous variances} \label{sec:unkown_het_var}

\begin{figure}[h] 
\centering
\includegraphics[width=0.7\textwidth]{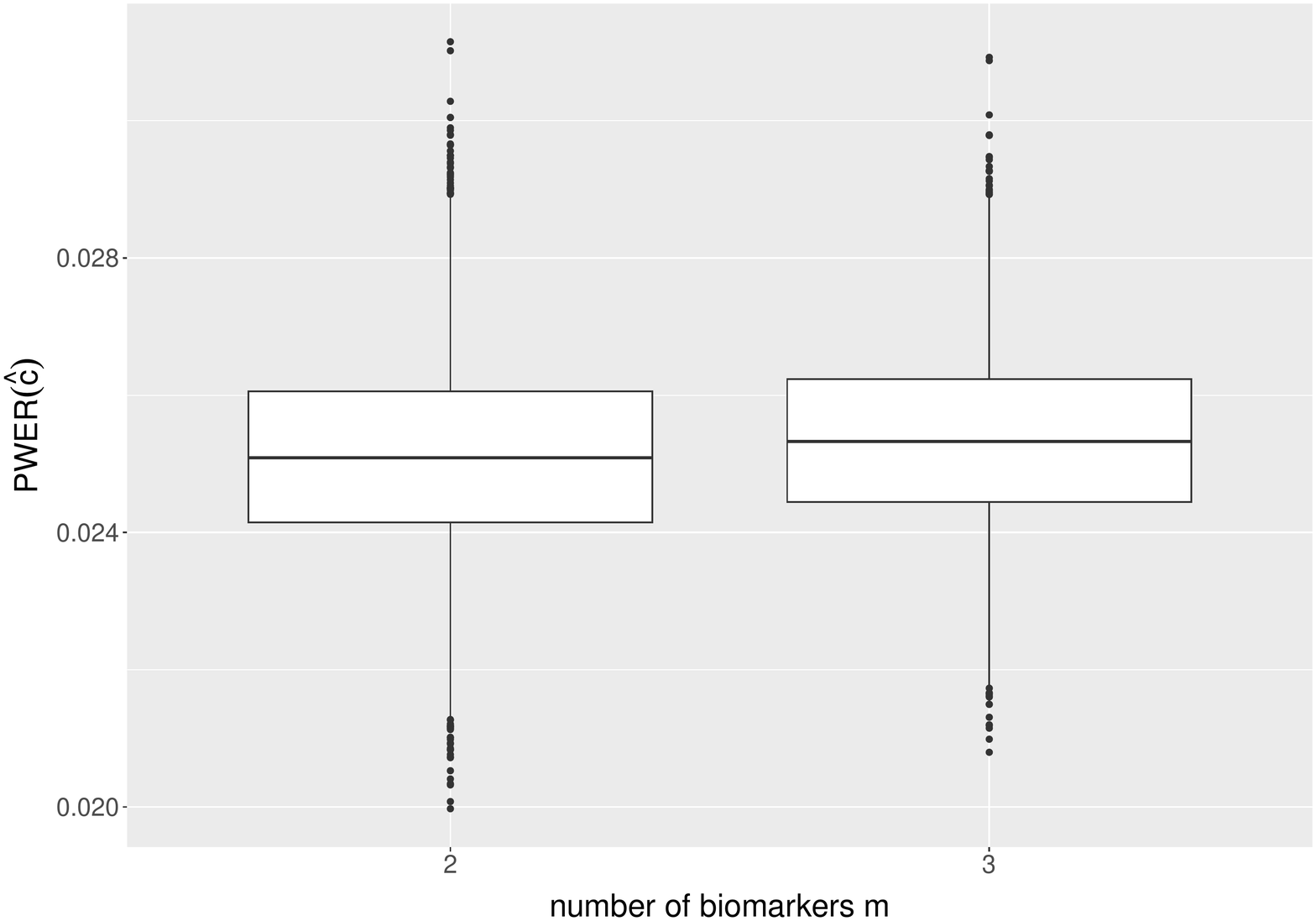}
\caption{Distribution of the approximated true population-wise error-rate (PWER) for heterogeneous residual variances between strata and treatments, for $m=2$ and $m=3$ populations.}
\label{fig5}
\end{figure}

Under the assumption of heterogeneous and unknown variances the estimated critical value cannot be determined from condition \ref{eq:pwer-est} anymore, because the joint distribution of the test statistics is unknown in this case. Instead we need to follow an approximate approach. For FWER-control, \cite{hasler} propose to determine population-specific critical boundaries $c_i$ from an $n$-dimensional $t$-distribution with degrees of freedom according to \cite{satterthwaite} \[df_i^*=\frac{\left(\hat{\sigma}_{\Pop_i, T_i}^2/n_{\Pop_i, T_i}+ \hat{\sigma}_{\Pop_i, C}^2/n_{\Pop_i, C}\right)^2}{\frac{\left(\hat{\sigma}_{\Pop_i, T_i}^2/n_{\Pop_i, T_i}\right)^2}{n_{\Pop_i, T_i}-1} + \frac{\left(\hat{\sigma}_{\Pop_i, C}^2/n_{\Pop_i, C}\right)^2}{n_{\Pop_i, C}-1}},\] and plug-in estimation of the correlation matrix. We use this distribution to find our estimated, population-specific critical values $\hat{c}_1, \dots, \hat{c}_n$ from 
\begin{align*} 
\sum_{J \subseteq I} \hat{\pi}_J \left(1- F_{\zero, \hat{\bSigma}_J, df_i^*}(\hat{c}_i, \dots, \hat{c}_i) \right) = \alpha \text{ for all } i \in I,
\end{align*} where $F_{\zero, \hat{\bSigma}_J, df_i^*}$ denotes the cdf of the $|J|$-dimensional $t$-distribution with parameters $\zero$, $\hat{\bSigma}_J = (\hat{\Sigma}_{ij})_{i, j \in J}$ and $df_i^*$ degrees of freedom. To approximate the resulting true PWER, we generate 10\,000 random samples under the global null hypothesis, store the test results for each sample, and take the weighted average over all strata-wise proportions of rejected hypotheses: \[\PWER(\hat{c}_1, \dots, \hat{c}_n) \approx \sum_{J \subseteq I} \pi_J \frac{\# \text{ reject at least one } H_j, j \in J}{\# \text{ simulation runs}}\] The results for the configuration presented before (with $N=500$ patients, pairwise different treatments, independent biomarkers, equal patient allocation and $\alpha = 0.025$) with randomly generated variances are contained in figure \ref{fig5} (only for $m=2$ and $m=3$ populations due to the increased computational effort needed). The corresponding summary statistics can be found in Appendix \ref{app:tables}. We see that the true PWER is again controlled very well in the average, but now observe a higher variance of the simulated values. This is due to the only approximate calculation of the true PWER. We have also tried this approximation for some cases from Section \ref{sec:results} (where the true PWER was known) and found a similar variance there.

\section{Discussion} \label{sec:disc}

The PWER is a new concept for measuring multiple type I errors in clinical trials with overlapping patient populations. It gives the average probability of erroneous rejections in the disjoint population strata. This makes the PWER only consider type I errors that are really relevant to the patients. The aim of this work was to investigate the stability of the PWER under plug-in estimation of the (usually unknown) strata prevalences. We have seen that in situations with up to eight different biomarkers and up to 255 strata, estimating prevalences does not prevent from adequate control of the true PWER. When interpreting the results from Section \ref{sec:sim}, one should distinguish between individual values of the true PWER, which are conditional on the vector of sample sizes $(n_J)_{J \subseteq I}$, and their mean over all simulation runs. An individual value of $\PWER(\hat{c})$ is only meaningful for a specific study with a specific vector of sample sizes drawn from the multinomial distribution. We have seen a rather limited fluctation of these values. The unconditional PWER, where we average over all sample size configurations, approximates the expected PWER over many studies. For critical boundaries that are based on sample estimates of the prevalences, it is very well under control. In practice the situation may even be improved by also utilizing historical data in the estimation of the prevalences. We were also able to prove that the true PWER converges almost surely towards the significance level $\alpha$, i.e., its asymptotic distribution is degenerated.

Regarding the maximal strata-wise FWER, we have seen in the simulations that under PWER control it is often indirectly controlled at a higher level, here $2\alpha$. Note that there is no guarantee for this, so in practice the maximal SWER should always be investigated (e.g.\ by simulations) and be taken into account in order to adjust the PWER or the prevalences used, as indicated in Section \ref{sec:min} (if deemed necessary). However, when using the same critical value for all test statistics, it can be expected that by PWER control the FWER is at least smaller than the FWER with no multiplicity control.

One might think that for high numbers of strata (e.g. 255 strata for $m=8$ populations) and low correlations, PWER-control would quickly lead to overly conservative tests. However, note that in the setting with overlapping populations, conservatism does not increase with the number of strata, but only with the number of populations that determines the number of tests to be performed. By formula \ref{eq:corr}, the correlations only depend on the relative size of the intersections of populations in the control group which does not necessarily decrease with increasing number of populations. For example, consider a sequence of independent binary biomarkers -- the proportion of a group with a specific set of multiple expressed biomarkers in its populations is then independent of the number of biomarkers (and just depends on the nature of these biomarkers).

An important prerequisite for the practical applicability of the PWER is the development of a sample size estimation for PWER-controlling studies, which is a subject for future reasearch. The convergence of the rejection boundaries presented in Appendix \ref{app:conv} could be of interest for this. Due to the restriction on type I errors that are relevant to the patients, we expect PWER-control to enable higher power and lower sample sizes than FWER-control.

\section*{Acknowledgements}

The authors thank the anonymous referees for their helpful comments. We would also like to thank Dr.\ Miriam Kesselmeier for kindly providing the data set of her real data example, and Dr.\ Charlie Hillner for his comments on a previous version of the manuscript.

\section*{Declaration of conflicting interests}

The author(s) declared no potential conflicts of interest with respect to the research, authorship, and/or publication of this article.

\appendix
\renewcommand{\thesubsection}{\Alph{subsection}}
\section*{Appendix}

\subsection{Proof of theorem \ref{thm1}} \label{app:proof_thm1}

For every $\btheta$ and $J \subseteq I$ with $J \cap I_0(\btheta) \neq \emptyset$ we have $\btheta \in H_{J \cap I_0(\btheta)}$. So according to the subset pivotality condition (\ref{eq:sp}) there is a $\btheta^* \in H_I$ with
$\FWER_{\btheta}^J \coloneqq  \P_{\btheta}\left(\max_{j \in J\cap I_0(\btheta)} Z_j > c \right) = \P_{\btheta^*}\left(\max_{j \in J\cap I_0(\btheta)} Z_j > c  \right) \leq \P_{\btheta^*}\left(\max_{j \in J} Z_j > c \right).$ With the monotonicity (\ref{eq:sm}) we get that $\FWER_{\btheta}^J \leq \P_{\zero}\left(\max_{j \in J} Z_j > c \right) = \FWER_{\zero}^J.$ One concludes $\PWER_{\btheta} = \sum_{J \subseteq I} \pi_J \FWER_{\btheta}^J \leq \sum_{J \subseteq I} \pi_J \FWER_{\zero}^J = \PWER_{\zero}$.

\subsection{Correlation of the test statistics} \label{app:corr}

We calculate the correlation matrix of the test statistics from section \ref{sec:knownvar}. First, we note that
\begin{align*}
V_i = \Var\left(\hat{\mu}_{i,T_i}-\hat{\mu}_{i,C}\right) &= \Var\left(\sum_{J \subseteq I:\, i \in J}\frac{n_{J, T_i}}{n_{i,T_i}}\bar{X}_{J,T_i} \right) + \Var\left(\sum_{J \subseteq I: \, i \in J} \frac{n_{J, C}}{n_{i,C}}\bar{X}_{J,C}\right) \\
&\phantom{:}= \sum_{J \subseteq I:\, i \in J} \left( \frac{n_{J,T_i}}{n_{i,T_i}}\right)^2 \Var(\bar{X}_{J,T_i}) + \sum_{J \subseteq I:\, i \in J} \left( \frac{n_{J,C}}{n_{i,C}}\right)^2 \Var(\bar{X}_{J,C}) \\
&\phantom{:}=  \sum_{J \subseteq I:\, i \in J} \frac{n_{J, T_i}}{n_{i,T_i}^2}\sigma_{J, T_i}^2 +  \frac{n_{J,C}}{n_{i,C}^2}\sigma_{J, C}^2,
\end{align*}
using the independence of the observations in the second equation. For different treatments tested in all populations and $i \neq j$ we have
\begin{align*}
\Corr(Z_i, Z_j) &= \frac{\Corr\left(\hat{\mu}_{i,T_i} - \hat{\mu}_{i,C}, \hat{\mu}_{j,T_j} - \hat{\mu}_{j,C}\right)}{\sqrt{V_iV_j}}= \frac{\Corr\left(\hat{\mu}_{i,C}, \hat{\mu}_{j,C} \right)}{\sqrt{V_iV_j}} ,
\end{align*}
because correlations with unequal treatments eliminate due to independent observations. And it holds
\begin{align*}
\Corr\left(\hat{\mu}_{i,C}, \hat{\mu}_{j,C} \right) &= \Corr\left( \sum_{J \subseteq I:\, i \in J} \frac{n_{J, C}}{n_{i,C}} \bar{X}_{J,C}, \sum_{J' \subseteq I:\, j \in J'} \frac{n_{J', C}}{n_{j,C}} \bar{X}_{J',C}\right) \\
&= \sum_{J \subseteq I:\, i \in J}\sum_{J' \subseteq I:\, j \in J'} \frac{n_{J,C} n_{J',C}}{n_{i,C} n_{j,C}} \Corr\left(\bar{X}_{J,C}, \bar{X}_{J',C}\right)\\
&= \sum_{J \subseteq I:\, i,j \in J} \frac{n_{J,C}^2}{n_{i,C} n_{j,C}} \text{Var}\left(\bar{X}_{J,C}\right) =\sum_{J \subseteq I:\, i,j \in J}  \frac{n_{J,C}}{n_{i,C} n_{j,C}}\sigma_{J,C}^2 . 
\end{align*}
All in all, this gives \[\Corr\left(Z_i, Z_j\right) = \frac{\sum_{J \subseteq I:\, i,j \in J} n_{J,C} \sigma_{J,C}^2}{n_{i,C} n_{j,C} \sqrt{V_iV_j}}.\]
In case of only one single treatment $T$ tested in all populations one obtains analogously \[\Corr\left(Z_i, Z_j\right) = \frac{1}{\sqrt{ V_iV_j}}\sum_{J \subseteq I:\, i,j \in J} \left(\frac{n_{J,T} \sigma_{J,T}^2}{n_{i,T}n_{j,T}} + \frac{n_{J,C} \sigma_{J,C}^2}{n_{i,C}n_{j,C}}\right).\]

\subsection{Convergence of the true PWER} \label{app:conv}

Under the conditions of section \ref{sec:unknown-hom-var} we assume that for every parameter $N \in \N$ of the multinomial distribution, and for each of its realizations $(n_J)_{J \subseteq I}$, $\hat{c}_N$ is a critical value that satisfies $\widehat{\PWER}(\hat{c}_N) = \alpha$. The sequence $\left(\PWER(\hat{c}_N)\right)_{N \in \N}$ then almost surely converges to $\alpha$. This follows from $\PWER(\hat{c}_N) - \alpha = \PWER(\hat{c}_N) - \widehat{\PWER}(\hat{c}_N) = \sum_{J \subseteq I} \left(\pi_J - \hat{\pi}_J\right)\left(1-F_{\mathbf{0}, \mathbf{\Sigma}_J}\left(\hat{c}_N, \dots, \hat{c}_N \right) \right) \xrightarrow[N \to \infty]{\text{a.s.}}0,$ in consequence of the strong consistency of the MLE $\boldsymbol{\hat{\pi}}$ and the boundedness of $F$.

The sequence $(\hat{c}_N)_{N \in \N}$ of the estimated critical values is also convergent, in terms of convergence in probability, towards the critical limit $c(\bpi)$ that results from using the true prevalences and correspondingly proportioned sample sizes. To prove this, we first note that for any fixed $c \in \R$ the estimated PWER $S_N(c) \coloneqq \widehat{\PWER}_{\boldsymbol{\nu}_N, \bSigma_N}(c)$ converges in probability (and even almost surely) to $S(c) \coloneqq \PWER_{\boldsymbol{\nu}(\bpi), \bSigma(\bpi)}(c)$, where $\boldsymbol{\nu}(\bpi)$ and $\bSigma(\bpi)$ are based on the true prevalences. This follows from the continuity of the multivariate t-distribution with respect to its location and scale matrix. Let $\varepsilon > 0$ be given. Using the strong monotonicity of the PWER with respect to $c$, we find
\begin{align*}
\P\left(\left|\hat{c}_N - c(\bpi)\right| \geq \varepsilon\right) &\leq \P\left(\hat{c}_N \geq \varepsilon +  c(\bpi) \right) + \P\left( \hat{c}_N \leq c(\bpi) - \varepsilon\right) \\
&\leq \P\left(S_N\left(\varepsilon +  c(\bpi)\right) \geq \alpha \right) + \P\left( S_N\left(c(\bpi) - \varepsilon\right) \leq \alpha\right) \\
&= \P\left(S_N\left(\varepsilon +  c(\bpi)\right) - S(\varepsilon + c(\bpi)) \geq \alpha - S(\varepsilon + c(\bpi)) \right) \\
& \quad+ \P\left( S(c(\bpi)-\varepsilon) -S_N\left(c(\bpi) - \varepsilon\right) \geq S(c(\bpi)-\varepsilon) - \alpha\right),
\end{align*}
and also some $\delta_1, \delta_2 > 0$ with $
S(\epsilon + c(\bpi)) = \alpha - \delta_1$ and $S(c(\bpi) - \varepsilon) = \alpha + \delta_2.$
Hence with the pointwise convergence of $S_N$ to $S$ we get
\begin{align*}
\P\left(\left|\hat{c}_N - c(\bpi)\right| \geq \varepsilon \right) 
&\leq \P\left(\left|S_N\left(\varepsilon +  c(\bpi)\right) - S(\varepsilon + c(\bpi))\right| \geq \delta_1 \right)\\
& \quad + \P\left( \left|S(c(\bpi)-\varepsilon) -S_N\left(c(\bpi) - \varepsilon\right)\right| \geq \delta_2\right) \xrightarrow[N \to \infty]{} 0.
\end{align*}

\newpage

\subsection{Summary statistics for the simulated values} \label{app:tables}

\centering
\begin{tabular}{llllllllll}
\hline
&$m$&$N$&Mean&SD&Min&Q1&Med&Q3&Max \\ \hline
&2&500&0.02500&0.00039&0.02272&0.02476&0.02500&0.02525&0.02680 \\
&3&500&0.02501&0.00042&0.02245&0.02474&0.02501&0.02528&0.02669 \\
\multirow{3}{*}{\makecell{true PWER \\ (figure \ref{fig:sim1})}}&4&500&0.02501&0.00041&0.02341&0.02474&0.02501&0.02527&0.02666 \\
&5&500&0.02501&0.00039&0.02294&0.02476&0.02500&0.02525&0.02684 \\
&6&500&0.02500&0.00037&0.02333&0.02477&0.02500&0.02523&0.02683 \\
&7&500&0.02501&0.00035&0.02341&0.02479&0.02500&0.02522&0.02679 \\
&8&500&0.02501&0.00033&0.02346&0.02480&0.02501&0.02521&0.02680 \\ \hline
&3&25&0.02516&0.00186&0.01597&0.02390&0.02506&0.02632&0.03337 \\
\multirow{4}{*}{\makecell{true PWER \\ (figure \ref{fig:sim2})}}&3&50&0.02508&0.00132&0.02009&0.02418&0.02504&0.02593&0.03044 \\
&3&100&0.02504&0.00095&0.02136&0.02441&0.02501&0.02563&0.02873 \\
&3&150&0.02503&0.00078&0.02110&0.02452&0.02501&0.02552&0.02879 \\
&3&200&0.02502&0.00067&0.01913&0.02459&0.02501&0.02544&0.02777 \\
&3&500&0.02501&0.00042&0.02245&0.02474&0.02501&0.02528&0.02669 \\ \hline
&2&500&0.03923&0.00601&0.02512&0.03461&0.03961&0.04423&0.04936 \\
&3&500&0.04597&0.00961&0.02563&0.03859&0.04519&0.05263&0.07310 \\
\multirow{3}{*}{\makecell{max SWER \\ (figure \ref{fig:sim3})}}&4&500&0.04894&0.01126&0.02637&0.04035&0.04740&0.05593&0.09563 \\
&5&500&0.05020&0.01193&0.02722&0.04132&0.04823&0.05706&0.10460 \\
&6&500&0.05068&0.01197&0.02702&0.04203&0.04840&0.05713&0.11332 \\
&7&500&0.05079&0.01161&0.02836&0.04242&0.04868&0.05669&0.11754 \\
&8&500&0.05070&0.01102&0.02896&0.04289&0.04866&0.05626&0.11470 \\ \hline
&2&500&0.02651&0.00392&0.01740&0.02348&0.02675&0.02978&0.03312 \\
&3&500&0.02651&0.00392&0.01740&0.02348&0.02675&0.02978&0.03312 \\
\multirow{3}{*}{\makecell{mean SWER \\ (figure \ref{fig:sim3})}}&4&500&0.02689&0.00601&0.01504&0.02230&0.02604&0.03060&0.05200 \\
&5&500&0.02684&0.00620&0.01511&0.02222&0.02579&0.03038&0.05543\\
&6&500&0.02678&0.00614&0.01491&0.02234&0.02558&0.03005&0.05939 \\
&7&500&0.02673&0.00592&0.01551&0.02246&0.02562&0.02970&0.06135 \\
&8&500&0.02665&0.00560&0.01582&0.02268&0.02560&0.02945&0.05970\\ \hline
\multirow{2}{*}{\makecell{true PWER \\ (figure \ref{fig5})}}&2&500&0.02512&0.00140&0.01997&0.02415&0.02509&0.02606&0.03115 \\
&3&500&0.02535&0.00134&0.02080&0.02445&0.02533&0.02623&0.03093 \\ \hline
\end{tabular}

\end{document}